# Reactor driven neutron sources from the $^{207}$Pb(γ,n) and $^{205}$Tl(γ,n) reactions

Sylvian Kahane [a], Yair Ben-Dov (Birenbaum) [b], Raymond Moreh [c,*]

[a] *P.O. Box 1630, 84965, Omer, Israel*

[b] *Nuclear Research Center, Negev, 84190, Beer-Sheva, Israel*

[c] *Department of Physics, Ben-Gurion University, 84120, Beer-Sheva, Israel*

**ABSTRACT**

Mono-energetic γ-beams (Δ ~ 10 eV) based on thermal neutron capture, in a nuclear reactor, using the V(n,γ) and Fe(n,γ) reactions were utilized for generating fast neutron sources from lead and thallium respectively, via the $^{207}$Pb(γ,n) and $^{205}$Tl(γ,n) reactions. It so happened that one of the incident γ-lines of the V source, $E_\gamma$ = 7163 keV, photoexcites by chance a resonance level in $^{207}$Pb, which emits neutrons at an energy of 423 keV. In a similar manner the incident γ-line at $E_\gamma$ = 7646 keV of the Fe(n,γ) source photoexcites by chance a resonance level in the $^{205}$Tl isotope, which emits neutrons at an energy of 99 keV. The cross sections for the neutron emission process were measured and found to be σ(γ,n) = 35±6 mb and 107±17 mb respectively with intensities of the order of $10^4$ n/sec.

- - - - - - - - - - - - - --

* Corresponding author

Email addresses: sylviankahane@gmail.com (S. Kahane), bendovori@gmail.com (Y. Ben-Dov), moreh@bgu.ac.il (R. Moreh)

## 1. Introduction

Neutrons sources can be produced, for research purposes, through a variety of one- and two-step nuclear reaction chains[1]. Nuclei that decay via α emission, such as $^{210}$Po, $^{239}$Pu, $^{226}$Ra, $^{241}$Am and $^{242}$Cm, can produce neutrons by mixing them with $^9$Be (i.e., using the $^9$Be(α,n) reaction). More rarely, mixtures with Boron, Fluorine or Carbon are used. This type of sources is prepared by mixing the $^9$Be with the α emitter[2-3] in a fine powder form, due to the short range of the α in the heavy nuclei. The most useful source of this type is the AmBe, which was offered for a long time by a majority of vendors and it is heavily used at a large number of labs. At this time, US has ceased its $^{241}$Am production and very few vendors[4], using Americium of Russian origin, are available.

Neutrons can also be produced with photonuclear reactions, where the initiating photons can come from either the decay of radioisotopes such as $^{124}$Sb (which emits γ-rays of energies higher than 1.665 MeV, being the threshold for n-emission in $^9$Be – i.e., using the $^9$Be(γ,n) reaction) or from bremsstrahlung produced by electron accelerators on high-Z targets. For example, at Gaertner Linear Accelerator Laboratory at RPI[5,6], a bremsstrahlung produced white neutron spectrum is converted into several discrete energies using filters of pure iron or $^{238}$U.

Accelerators can also be used to provide charged particles (protons or deuterons, primarily) for either nuclear reaction-based neutron sources, such as $^7$Li(p,n)$^7$Be reaction (generating neutrons in the 0 to 1.3 MeV range[7]), or spallation based neutron sources like ISIS[8], SNS[9], and n_TOF[10,11]. Finally, neutrons can be produced by fission reactions, either in critical assemblies or via nuclei which decay by spontaneous fission, such as $^{252}$Cf[1].

In this work, we present a method of producing neutron sources based on the (γ,n) reactions on various elements using high energy photons of 7 to 10 MeV originated in the thermal n-capture in a nuclear reactor[12]. Therefore, such neutron sources are reactor driven and require three stringent conditions for their production to be feasible: (1) one of the incident γ-lines produced by the (n,γ) source should resonantly photoexcite, *by chance, an isolated* nuclear level in the target (sample); (2) the photoexcited level should be an *unbound* level higher than the threshold energy for neutron emission; (3) the cross section for n-emission should be high enough for producing a neutron source of relatively high intensity. The obtained intensities are of the order of $10^3 - 10^5$ n/sec, comparable to those of the (α,$^9$Be) type or $^{252}$Cf (see Ref. 4 specifically for $^{241}$AmBe intensities available).

Here we deal with two neutron sources: The first is based on the V(n,γ) reaction where the 7.163 MeV gamma ray is generated by thermal neutron capture in $^{51}$V with subsequent deexcitation to the 147.8 keV level of $^{52}$V. This γ line happens to photoexcite resonantly an unbound level in $^{207}$Pb inducing a $^{207}$Pb(γ, n)$^{206}$Pb reaction which leaves the residual $^{206}$Pb nucleus (Fig. 4) at its ground state with the emission of the 423 keV neutron group (there is a small spread in energy due to the kinematics). The second source is similar, thermal neutron capture in $^{56}$Fe produces a 7.646 MeV γ with the resulting $^{57}$Fe at the ground state. $^{205}$Tl is resonantly photoexcited by this gamma, emits neutrons at $E_n$ =99 keV while the residual $^{204}$Tl is left at the ground state (Fig. 7). These neutron sources are driven by an operating nuclear reactor which creates photons via the V(n,γ) or Fe(n,γ) reactions followed by neutrons produced via the (γ,n) reaction. Such accidental photoexcitation processes are not very rare[13-16]. This may be understood if one considers the large number of the discrete γ lines in each photon beam and also the density of nuclear levels occurring in each nucleus at excitations of 7 to 10 MeV. Both the incident γ line and the nuclear level are Doppler broadened having Δ ~ 10 eV where the broadening depends on the energy and the nuclear mass, as explained in the Appendix. In the past, similar studies of neutron emitting resonance levels (by photoexcitation of isolated *resonance* levels) were reported in Ref. 13-16. In Ref. 13 a γ line at 7632 keV of the Fe(n,γ) reaction photoexcites one nuclear level in $^{207}$Pb producing a strong neutron source having $E_n$ = 86 keV. In Ref. 14 the γ line at 7637 keV of the Cu(n,γ) reaction overlaps a nuclear level in $^{209}$Bi emitting strong intensity neutron groups at $E_n$ = 114 keV and 177 keV. In Ref. 15, a γ-line at 8884 keV of the Cr(n,γ) reaction photoexcites a level in $^{49}$Ti emitting neutrons at $E_n$ = 726 keV. Finally in Ref. 16 $^{208}$Pb was photoexcited by the 7632 keV γ-line of the Fe(n,γ) reaction emitting neutrons at $E_n$ = 262 keV. Table 1 lists a summary of the properties of all neutron sources reported in the present work together with all previously reported sources (Refs. 13-16) generated using the same method.

## II. Experimental Method

The experimental system is shown schematically in Fig. 1 where the γ source is produced by either the V(n,γ) reaction or the Fe(n,γ) reaction. The lifetime of the compound nucleus produced by thermal neutron capture is of the order of $10^{-14}$ sec, i.e., it decays promptly emitting γ- rays (in the case of bound levels) or both γ-rays and neutrons (for unbound levels).

**Table I. Summary of resonantly produced quasi monoenergetic neutron groups via the (n,γ) + (γ,n) chain reactions.**

| n-group energy [keV] | σ(γ,n) [mb] | Intensity at our installation [n/sec] | Primary (n,γ) source (isotope) | γ energy [MeV] which excites resonantly a secondary target | Secondary (γ,n) source of neutrons (isotope) | |
|---|---|---|---|---|---|---|
| 423 | 35±6 | 2.0x10$^3$ | $^{51}$V | 7.163 | $^{207}$Pb | Present work |
| 99 | 107±17 | 4.0x10$^4$ | $^{56}$Fe | 7.646 | $^{205}$Tl | Present work |
| 86 | 370±50 | 1.0x10$^5$ | $^{56}$Fe | 7.632 | $^{207}$Pb | Ref. 13 |
| 114 and 177 | 93±14 275±40 | 1.0x10$^5$ (combined) | Cu | 7.637 | $^{209}$Bi | Ref. 14 |
| 726 | 33±4 | Not given | Cr | 8.884 | $^{49}$Ti | Ref. 15 |
| 262 | 50±7 | Not given | $^{56}$Fe | 7.632 | $^{208}$Pb | Ref. 16 |

Such γ sources are of huge intensities, being produced using kilogram amounts of metal mounted in tangential beam tubes, near but outside the core of the Israel Research Reactor 2 (IRR-2). The typical thermal neutron flux, in the tubes, at this reactor, is of the order ~ 10$^{12}$ n/cm$^2$/s (based on internal calibrations performed at IRR-2 from time to time). For a $^{51}$V mass of 1.8 Kg (see below), with a thermal capture cross section[17] of 4.9 b, a total production γ rate of ~8x10$^{13}$ photons/sec is predicted in 4π, not accounting for self-absorption. The γ intensities in the strong individual lines are expected to be one or two order of magnitude lower.

The V(n,γ) source was in the form of 6 separated metallic discs each 1 cm thick and 8 cm diameter with a spacing of 2 cm from each other. The corresponding mass of $^{51}$V is 1.8 Kg (99.76%). The resulting γ beam was collimated and neutron filtered yielding intensities of ~10$^6$ photons/cm$^2$/sec (for the strong γ lines) at the target position. The distance between the (n,γ) source and the target position is ~ 6 m, hence the source strong γ lines rate is of the order of 4.5x10$^{12}$ photons/sec, as expected more than one order of magnitude lower compared with the total γ intensity estimated above.

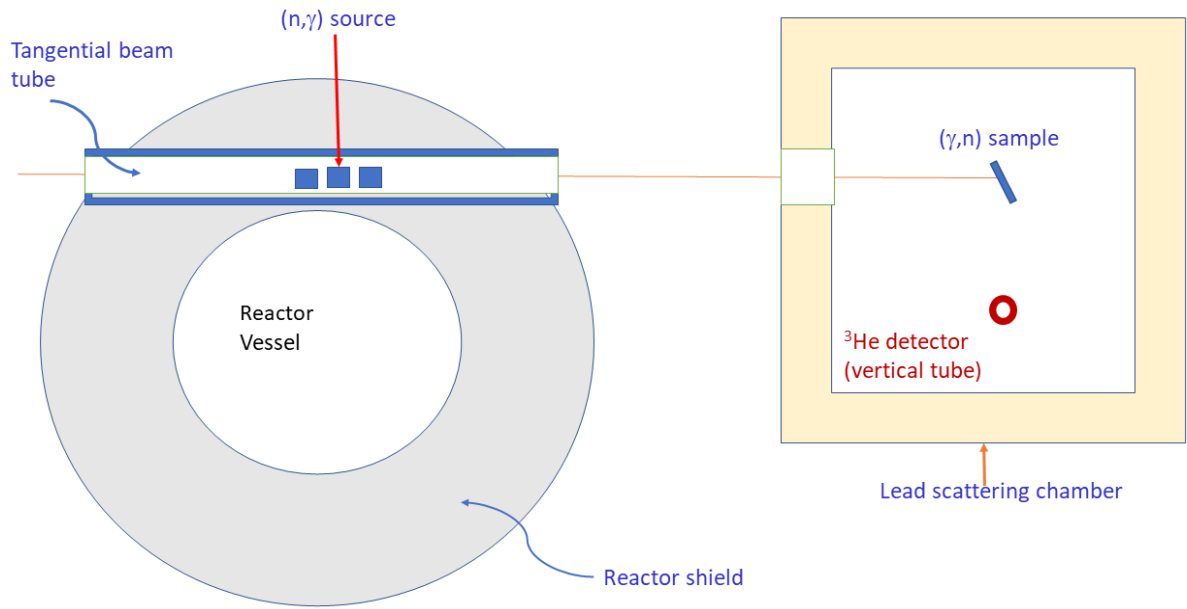

Fig. 1. Schematic diagram of the experimental setup (not to scale) showing the (n,γ) source set inside a beam tube tangential to the reactor core. The high intensity γ-beam was neutron filtered using a 40 cm borated parrafin absorber (for reducing the neutron background inside the scattering chamber). The scattering chamber is surrounded by walls of a 15 cm thick lead shielding.

The $^3$He detector, placed at 27 cm from the target, is a commercial neutron spectrometer manufactured by Seforad-Applied Radiation Ltd., Emek Hayarden, Israel, based on the gridded ionization counter of Shalev and Cuttler[18]. This cylindrical detector, 5 cm diameter, 15 cm active height, 1.2 mm thick (steel), is filled at 6 atm with $^3$He, 3 atm with argon and 0.5 atm with methane. Neutron detection relies on the $^3$He(n,p)T reaction where Q = 764 keV. For *thermal neutrons* the absorption cross section is very large, 5300 b, producing a strong peak at 764 keV in the neutron Pulse Height Spectrum (PHS). This peak corresponds to the energy sum of the emitted p and T($^3$H) ions deposited in the ionization chamber. As an illustration, we show in Fig. 2 a PHS taken from a *previous* measurement[13]:

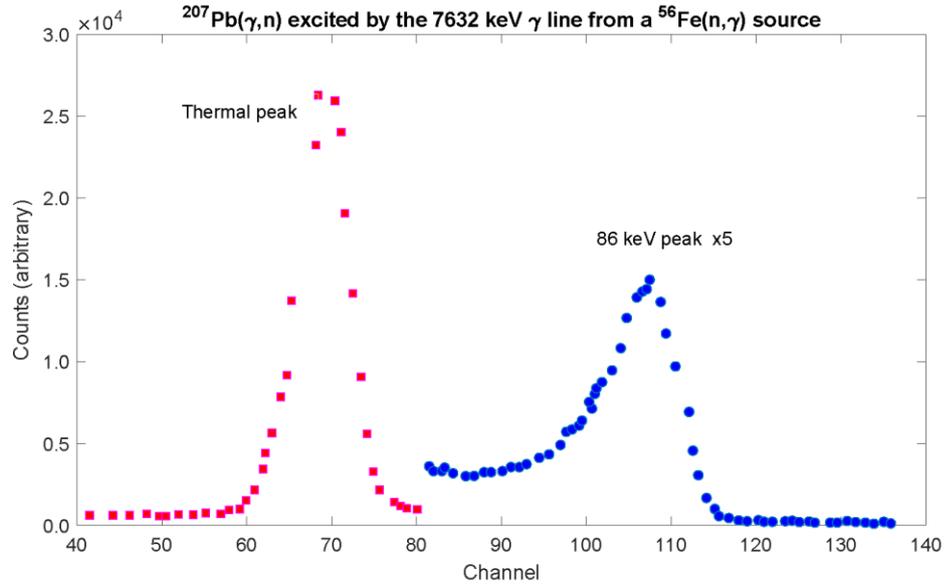

**Fig. 2. The thermal peak, i.e., the 734 keV energy deposition from the $^3$He(p,n)T reaction, in relation to another neutron group contribution**.

In the present work, the thermal peak is omitted from the figures and the energy scale of the fast neutrons is selected in such a way that its *zero value* starts at the peak of thermal neutrons. Details of energy calibration are described in Ref.13.

The signals from the $^3$He n-detector, were fed through a pre-amplifier to a shaping main amplifier producing gaussian shaped signals. The pulses from the amplifier were sorted using a Canberra 80 Analyzer where the obtained pulse height spectra could be analyzed for peak area calculation and energy calibration. The best energy resolution was obtained by operating the amplifier with a time constant τ ~ 12.8 μs. It was beneficial to increase τ with increasing neutron energy to allow for a better charge collection of the $^3$He detector. During the measurements the observed counting rate was lower than about 3000 cps, mainly due to the good collimation of the incoming gamma beam and to the target detector distance. Hence, no serious pileup effects were observed.

It should be emphasized that the use of a *tangential* beam tube is of paramount importance and one should avoid using a *radial* beam tube for producing the gamma source. This is because in a *radial* beam the amount of gamma and neutron background emerging from the reactor core is so huge that it overwhelms any gamma signal emitted by the (n, γ) reactions. At energies higher than ~ 5 MeV, this γ background consists mainly of *pileups* and hence cannot produce any resonance photoneutrons from the nuclei of the samples. Note also that such high intensity background photons can be elastically and inelastically scattered from the target and could block the $^3$He detector, being sensitive to γ-s also. The same is true regarding the neutron background coming from the radial tube.

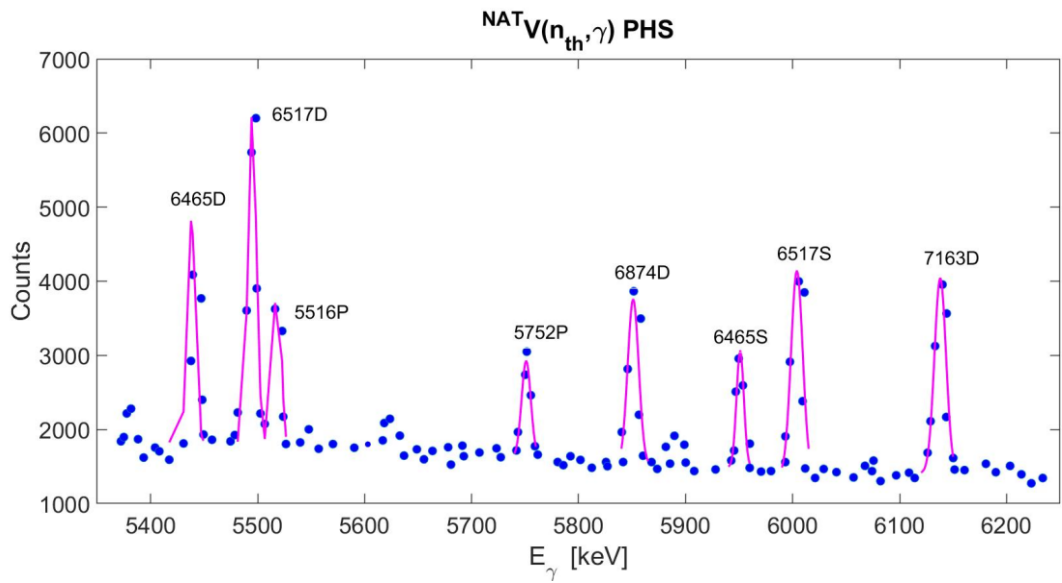

**Fig. 3.** High energy part of V(n$_{th}$,γ) pulse height spectrum (PHS), measured using a relatively small Ge detector (40cc), hence the double escape peaks (noted by D) have the strongest intensities (S = single escape, P = photopeak). The measurement was carried out by mounting the Ge detector at the position of the (γ,n) sample and inserting a 20 cm thick Pb attenuator for reducing the γ-line intensities to levels tolerated by the Ge detector. The strong intensity V γ-lines are emphasized by gaussians fitted to the measured points.

Fig. 3 shows the high energy part of the $^{nat}$V(n$_{th}$,γ) spectrum in the energy range 4 to 7.5 MeV, as measured with a 40 cc Ge detector in conjunction with a multichannel analyzer. The direct γ-beam was passed through a 40 cm borated parrafin absorber (not shown in Fig. 1); located along the tangential beam tube, for reducing the neutron background inside the scattering chamber. A resonance neutron emission via the (γ,n) reaction occurs only when one of the strong intensity γ lines such as the one at 7163 keV of Fig. 3 happens to overlap by chance and photoexcite an unbound nuclear level in $^{207}$Pb thus emitting 423 keV neutrons as discussed in more detail below.

### III. Results

## III.A. The 423 keV neutron source

As said, only the 7163 keV γ-line, originating from the $^{51}$V(n,γ) reaction, photoexcites resonantly a nuclear level in $^{207}$Pb. This level is unbound with a neutron separation energy of 6738 keV[19], decaying by neutrons of energy $E_n = 423$ keV, proceeding to the ground state of $^{206}$Pb. The neutron generation process is described in Fig. 4 and the corresponding pulse height spectrum (PHS), as measured using the high resolution $^3$He detector, is shown in Fig. 5; the detector system has a resolution of 17 keV for thermal neutrons and 24 keV for 1 MeV neutrons. The neutron group appearing at 620 keV is due to a chromium source, present in the same tangential beam tube, after the vanadium source (for experiments performed at the other end). In principle, this n-group could be avoided by removing the chromium from the beam tube. Experimentally, we could not make any changes because the (n,γ) source is highly radioactive after being irradiated for many years inside the tangential beam tube, thus the removal of the Cr γ-source is quite hazardous. The detector was shielded by wrapping it with 0.5 mm of metallic Cd and 2 mm of metallic Pb.

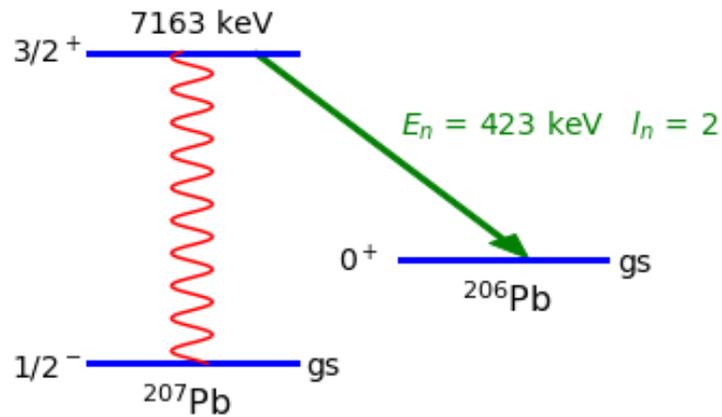

**Fig. 4.** Photoexcitation of the 7163 keV resonance level in $^{207}$Pb showing the photo absorption process and the subsequent 423 keV neutron emission (with $l_n$ =2) leading to the $^{206}$Pb ground state, $J_f^\pi = 0^+$, via the $^{207}$Pb (γ,n) reaction.

In this measurement the $^{207}$Pb enriched target ($^{207}$Pb 92.4%, 5.5% $^{208}$Pb, $^{206}$Pb 2.1%) was a 11.17 g square shaped (3.5 cm) in the form of powdered PbCO$_3$ containing a net weight of 8.0 g of $^{207}$Pb; its plane was inclined at 45° relative to the incident γ-beam and enclosed inside a very thin plastic film.

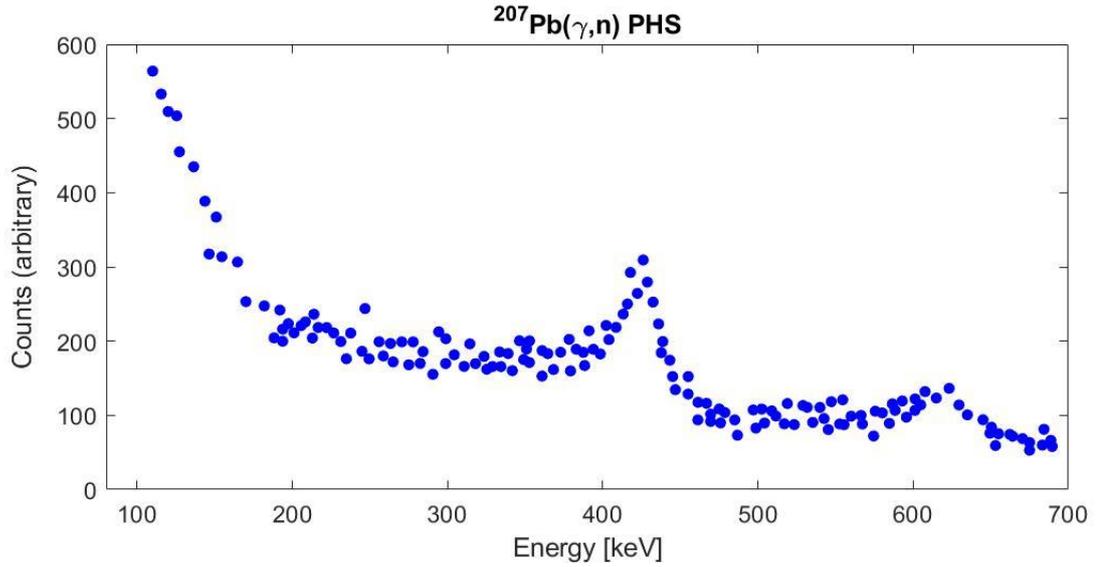

**Fig. 5.** Pulse height neutron spectrum of the $^{207}$Pb(γ,n) reaction measured with the $^{3}$He detector. The neutron peak is at an energy of 423 keV. The n-group at 620 keV is due to a contamination of the vanadium source with metallic chromium known to be present in the tangential beam port. Channel zero corresponds to the energy of the thermal neutron peak (see text).

The intensity of the produced neutron group is influenced by a series of local parameters like the weight and shape of the $^{51}$V γ-source, the Pb sample and the reactor flux etc. One parameter, of major importance is the σ(γ,n) cross section in the Pb target. In the present work, we are interested to quantify the total cross section σ (γ,n) of the resonant photoexcitation process. For this: 1) we determined the differential cross sections dσ/dΩ, at a scattering angle $\theta$=90° (see Fig. 1), and 2) we measured the angular distribution of the photoneutrons. These two quantities provided the reported cross sections. In addition, we also deduced other quantities of physical interest, such as the multipolarity of the emitted neutrons and the nuclear spin of the emitting level (see Appendix).

The dσ/dΩ differential cross section at $\theta$=90°, was measured relative to that of the 86 keV n-group (measured absolutely in a previous work[13]), by accounting for the relative efficiencies and the differences in the number of nuclei in the samples (otherwise the setups being identical), using the following relation:

$$\left(\frac{d\sigma}{d\Omega}\right)_{En} = \frac{Y_\theta^{En}}{Y_\theta^{86}} \frac{\mathcal{N}_{86}}{\mathcal{N}_{En}} \frac{\epsilon_n^{En}}{\epsilon_n^{86}} \left(\frac{d\sigma}{d\Omega}\right)_{86} \qquad (1)$$

Where $Y$ are the measured yields normalized per unit time, at 86 keV and at a neutron energy $E_n$ respectively, $\mathcal{N}$ is the number of nuclei in the two samples (respectively). Franz et. al.[21] used an identical detector and monoenergetic neutrons produced with a proton accelerator and the $^7$Li(p,n)$^7$Be reaction, for measuring the relative efficiencies of the detector in the 0.02 – 2.77 MeV range. The number of nuclei ratio is relevant only for the $^{205}$Tl target case and was found to be 1.078.

The measured angular distribution is shown in Fig. 6 at 5 angles between 50° and 130°; it was found to be symmetric around 90° which is characteristic of n-emission from a resonance level[15,16,20]. Some additional information on the angular distribution is given in the Appendix.

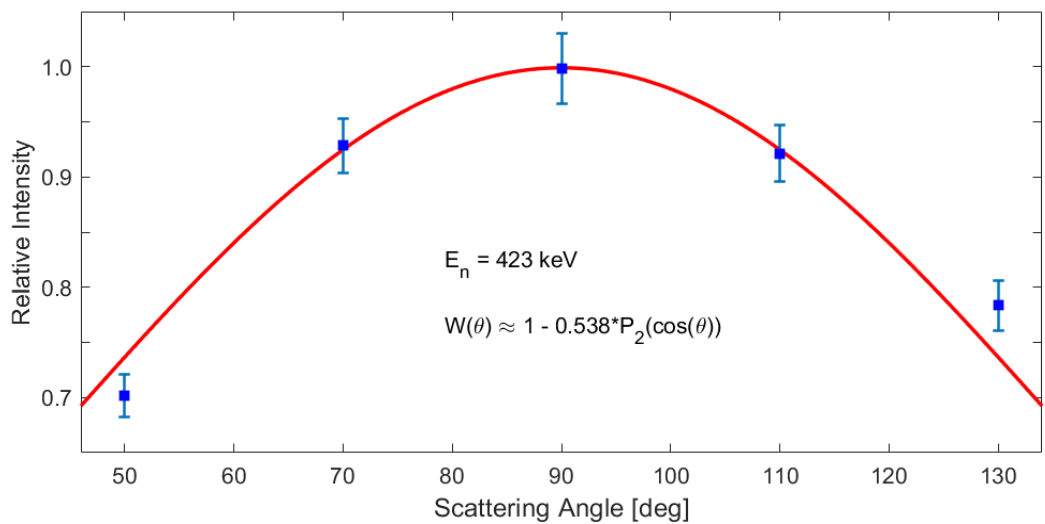

**Fig. 6. Angular distribution of the resonance neutron group at 423 keV. The error in the A2 parameter is about 10%.**

Thus, the cross section for the 7163 keV resonance in $^{207}$Pb was found to be $\sigma(\gamma,n) = 35 \pm 6$ mb. In our setup, the intensity of the 423 keV n-group was ~$2.0 \times 10^3$ n/sec.

Due to the recoil of the final nucleus ($^{206}$Pb), the energy of the emitted neutrons from the $^{207}$Pb($\gamma$,n) reaction varies with the n-emission angle relative to the incident γ-beam. The energy spread of the 423 keV neutrons reaching the $^3$He n-detector at 90° was calculated using Eq. 2 of Ref. 13; the angular opening subtended by the detector (at $\theta$=90°) is ± 6° and the corresponding energy spread was ± 80 eV.

***III.B. The 99 keV neutron source***

10identical detector and monoenergetic neutrons produced with a proton accelerator and the $^7$Li(p,n)$^7$Be reaction, for measuring the relative efficiencies of the detector in the 0.02 – 2.77 MeV range. The number of nuclei ratio is relevant only for the $^{205}$Tl target case and was found to be 1.078.

The measured angular distribution is shown in Fig. 6 at 5 angles between 50° and 130°; it was found to be symmetric around 90° which is characteristic of n-emission from a resonance level[15,16,20]. Some additional information on the angular distribution is given in the Appendix.

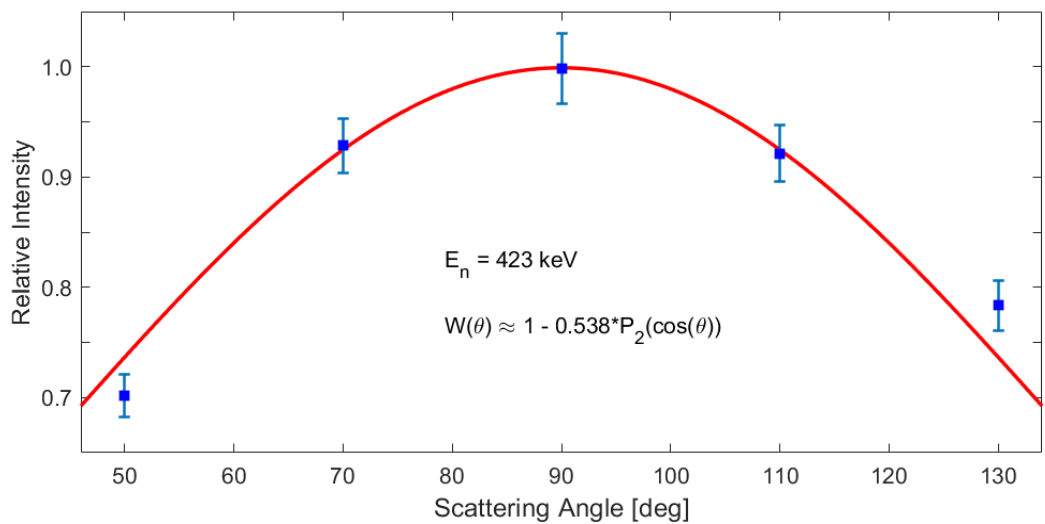

**Fig. 6. Angular distribution of the resonance neutron group at 423 keV. The error in the A2 parameter is about 10%.**

Thus, the cross section for the 7163 keV resonance in $^{207}$Pb was found to be $\sigma(\gamma,n) = 35 \pm 6$ mb. In our setup, the intensity of the 423 keV n-group was ~$2.0 \times 10^3$ n/sec.

Due to the recoil of the final nucleus ($^{206}$Pb), the energy of the emitted neutrons from the $^{207}$Pb($\gamma$,n) reaction varies with the n-emission angle relative to the incident γ-beam. The energy spread of the 423 keV neutrons reaching the $^3$He n-detector at 90° was calculated using Eq. 2 of Ref. 13; the angular opening subtended by the detector (at $\theta$=90°) is ± 6° and the corresponding energy spread was ± 80 eV.

***III.B. The 99 keV neutron source***

10identical detector and monoenergetic neutrons produced with a proton accelerator and the $^7$Li(p,n)$^7$Be reaction, for measuring the relative efficiencies of the detector in the 0.02 – 2.77 MeV range. The number of nuclei ratio is relevant only for the $^{205}$Tl target case and was found to be 1.078.

The measured angular distribution is shown in Fig. 6 at 5 angles between 50° and 130°; it was found to be symmetric around 90° which is characteristic of n-emission from a resonance level[15,16,20]. Some additional information on the angular distribution is given in the Appendix.

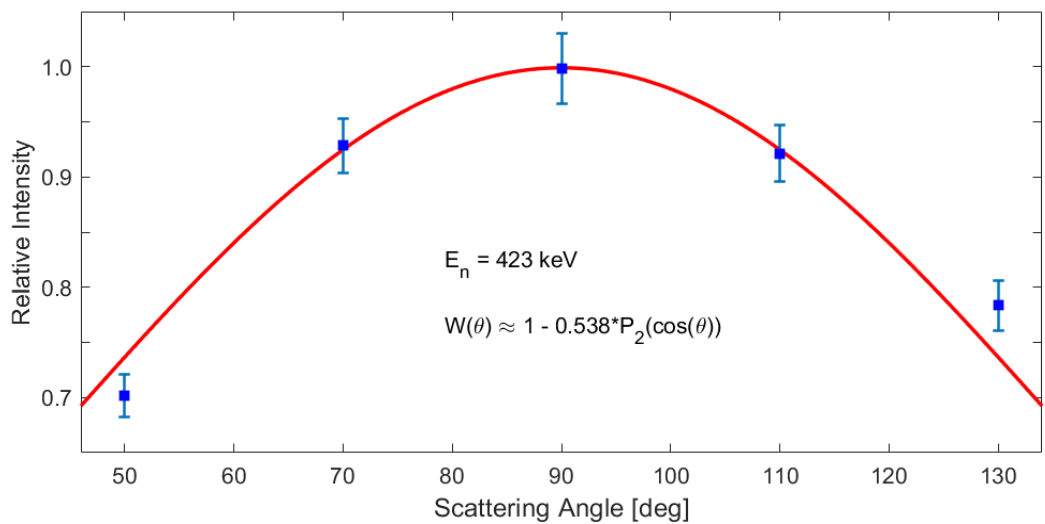

**Fig. 6. Angular distribution of the resonance neutron group at 423 keV. The error in the A2 parameter is about 10%.**

Thus, the cross section for the 7163 keV resonance in $^{207}$Pb was found to be $\sigma(\gamma,n) = 35 \pm 6$ mb. In our setup, the intensity of the 423 keV n-group was ~$2.0 \times 10^3$ n/sec.

Due to the recoil of the final nucleus ($^{206}$Pb), the energy of the emitted neutrons from the $^{207}$Pb($\gamma$,n) reaction varies with the n-emission angle relative to the incident γ-beam. The energy spread of the 423 keV neutrons reaching the $^3$He n-detector at 90° was calculated using Eq. 2 of Ref. 13; the angular opening subtended by the detector (at $\theta$=90°) is ± 6° and the corresponding energy spread was ± 80 eV.

***III.B. The 99 keV neutron source***

The iron γ-source was in the form of 5 metallic discs each 2 cm thick and 8 cm diameter with a spacing of 2 cm from each other. The Fe discs were placed in a separate tangential beam tube and near the core of the IRR-2 reactor. The generation process of this neutron source based on the Fe(n,γ) is described in Fig. 7 where the neutron emitting isotope is $^{205}$Tl. A 10.4 g disk shaped metallic Tl sample of 4.0 cm diameter was used, inclined at 45 deg relative to the incident gamma beam. Natural Tl consists of two isotopes $^{205}$Tl (70.5%) and $^{203}$Tl (29.5%). Only $^{205}$Tl was shown to resonantly scatter the 7646 keV γ line of the Fe(n,γ) reaction (Ref. 22). This level in $^{205}$Tl is unbound and is known to decay by two channels: photons and neutrons. The photon decay proceeds elastically to the ground state in $^{205}$Tl and inelastically to excited states as reported in detail in Ref. 22. This level also emits 99 keV neutrons leading to the ground state of $^{204}$Tl. In Fig. 8 we show a small part of the neutron PHS around 99 keV, as measured using the $^{3}$He detector. The entire spectrum (not shown) contains several other small peaks corresponding to other (γ,n) excitations arising from the other higher energy lines of the V(n,γ) source on $^{207}$Pb and on the other Pb isotopes of the sample as may be seen in Fig. 2 of Ref. 13.

The angular distribution of the resonantly scattered 7646 keV *photons* was measured previously and found to be isotropic. The measured cross section for the 7646 keV resonance in $^{205}$Tl is σ(γ,n) = 107±17 mb, and its intensity, in our setup, ~ 4.0x10$^4$ n/sec.

The total energy difference, between the forward to backward emitted neutrons, was 1.1 keV. At 90° the energy spread is ± 39 eV, calculated as in the case of the 423 keV group.

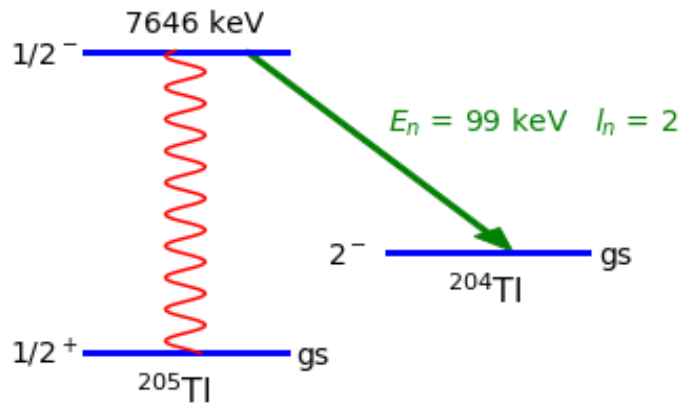

**Fig. 7. Photoexcitation of the 7646 keV resonance level in $^{205}$Tl showing the photo absorption process and the subsequent 99 keV neutron emission (with $l_n$ =2) leading to the $^{204}$Tl ground state, $J_f^\pi$ = 2$^-$, via the $^{205}$Tl(γ,n) reaction.**

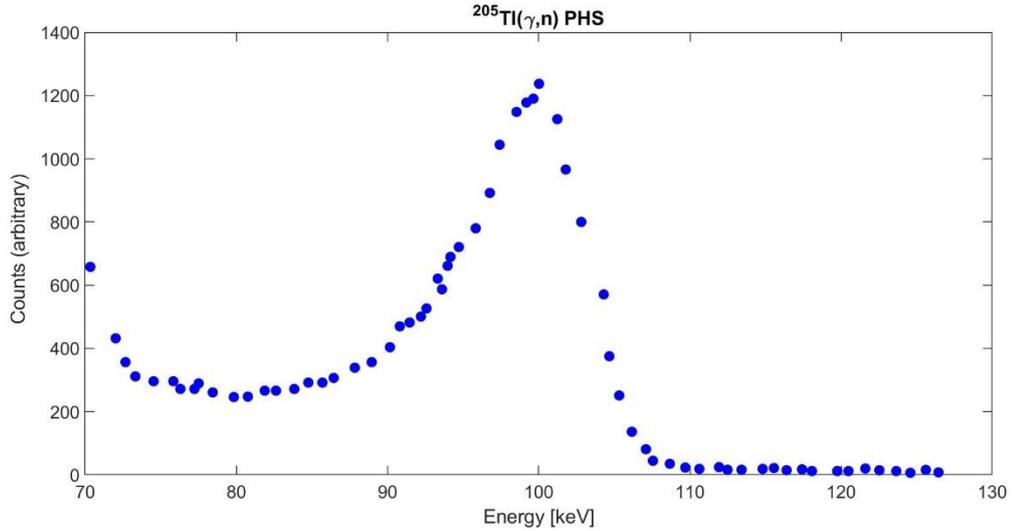

**Fig. 8.** Pulse height neutron spectrum emitted by the 7646 keV photoexcited level in $^{205}$Tl as measured using the $^3$He detector. The neutron peak is at an energy of 99 keV.

**IV. Background**

*IV.A. Neutron background*

We hereby deal with the neutron and gamma background in the scattering chamber. In principle there are two neutron background components: 1) Neutrons entering the scattering chamber directly from the beamline or originated in the beamline but scattered by collimators or shielding and structural materials. Here one can add also photoneutrons produced in these materials by the gammas present in the beam line. 2) Neutrons produced resonantly in the sample and subsequently bouncing from the walls, or photoneutrons produced both in the sample and in the surrounding walls.

The n-background reaching the scattering chamber directly from the beamline or from the walls of the tangential beam tube during reactor operation (component no. 1 above) is small, due to the use of the 40 cm long borated paraffin which is a very effective neutron shield. We routinely measured the neutron background in the absence of the resonant target. This was found to be small, some ~7% of the peak area in the case of the stronger 99 keV n-group. In the case of the 423 keV the situation is not as good, but a linear background can be reasonably subtracted (see Fig. 10).

The direct beam of the vanadium capture $\gamma$-rays consists of strong intensity lines at 6874, 7163, 7311 keV which are higher than the ($\gamma$,n) threshold of $^{207}$Pb (6738 keV). Background neutrons within the scattering chamber can be created mainly via the photonuclear ($\gamma$,n) reaction by these $\gamma$-lines on the sample (negligible because of the low weight of the samples and the *low cross section* from the tail of the giant dipole resonance) and/or on the lead walls. Such emitted neutrons can increase the $\gamma$-background via a subsequent (n,$\gamma$) reaction on the structural materials of the detector and of the lead shielding surrounding the detector, but this is a two step process with much lower probability.

*IV.B. Neutron multiple scattering*

Neutrons can reach the detector directly from sample (emitted neutrons of interest, not scattered) or scattered from the walls, or produced in the walls by photonuclear reactions. Here we address specifically the neutrons reaching the detector after one or more scatterings. The neutron multiple scattering (MS) from the lead walls of the scattering chamber was evaluated by a Monte Carlo simulation (using the MCNP code) for a 0.4 MeV neutron beam. The scattering chamber is modeled as four lead walls and a lead roof, each 15 cm thick. This construction stays on a steel pedestal 5 cm thick. The inside dimensions are 120 cm in each direction. The detector model is a $\phi$=5cm x h=15cm stainless cylinder of 1.2 mm thickness, filled with $^3$He gas at 6 atm, and covered with 2 mm lead and 0.5 mm cadmium. The neutron source is located along the $\gamma$-beam (see Fig. 1). The neutrons are emitted with the angular distribution given in Fig. 5. The detector is positioned at 27 cm from the source at 90° relative to the $\gamma$-beam direction (as in Fig. 1). The neutrons are monitored in the detector volume, not counting the neutrons arriving directly from the source with no scattering (MCNP is giving this quantity as a "flux" in units of 1/cm$^2$). This flux is subsequently multiplied by the $^3$He(n,p) cross section, by the density of the $^3$He gas (6 atm) and by the detector active volume giving effectively the number of occurring (n,p) reactions. The assumption is that all of them deposit energy and are counted (neglecting loses like the wall effect) providing the signal of the detector. This result is normalized by the MCNP code, to one neutron at source, hence the result is the relative fraction of 0.4 MeV produced in the sample and *counted* in the detector after MS in the walls. The above procedure is strictly accurate if the detector efficiency is exactly related to the (n,p) cross section. According to Franz[21], this is the case only under 100 keV. Above this energy one should multiply by the absolute efficiency and not by the (n,p) cross section. Unfortunately, Franz et. al. gave only the relative efficiency. Hence, we devised the following procedure: a) the flux was multiplied by the relative efficiency; b) in the region $E_n < 100$ keV a normalization factor was sought such that the results obtained with the (n,p) cross section are recovered; c) this normalization factor was used for all the energy range up to 400 keV. The curve based on the (n,p) cross section gives 50% more total MS (at 400 keV) compared with the normalized curve based on the relative efficiency. This difference is expected because the efficiency is lower than the cross section (Fig. 7 in Franz[21] et. al.). With increased neutron energy, the produced protons and tritons also have higher energies and their path, in the detector gas, is longer. Hence, they can escape the detector volume without a full deposition of their energies. The final result, given in Fig. 9, shows that there is a fraction of about $5.3 \times 10^{-7}$ (in a 15 keV bin) MS neutrons background at the nominal energy of the primary neutron energy, and a total of $8.3 \times 10^{-7}$ for all the MS neutrons. A similar simulation for the 100 keV neutrons (based only on the (n,p) cross section), gives a factor of 3 higher fractions of $2.1 \times 10^{-6}$ and $2.8 \times 10^{-6}$ respectively.

We estimated, by a separate simulation, that a flux of $10^5$ neutrons/cm$^2$/sec will enter the scattering chamber from the reactor. Combined with a hole of radius 10 cm in the scattering chamber wall close to the reactor, the neutron intensity is $3.1 \times 10^7$ n/sec. Using the above fractions, the background rate is expected to be ~25-75 n/sec, if the energy is in the 100-400 keV range. Using the above fractions is questionable because the majority of these neutrons, coming from the tangential beam tube outside the core, and possible after a number of scatterings in the collimator or the paraffin filter, are at thermal or low epithermal energy. According to the trend above between 0.4 and 0.1 MeV, higher fractions and higher rates are expected, but, and this is very important, due to the cadmium shield, the majority of these neutrons are not counted at all (according to Shalev and Cuttler[18], 0.4 mm of cadmium eliminates completely all the thermal neutrons). Even if their elimination is not 100% efficient, they are well below the energies of interest in the present work and not expected to influence the results for the two neutron sources.

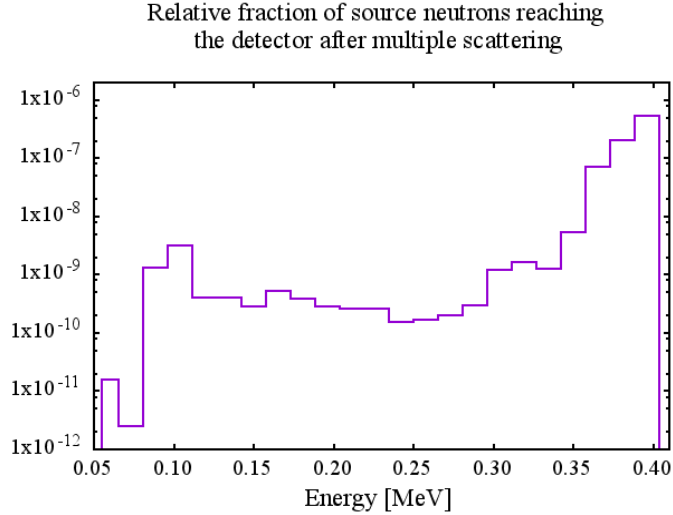

**Fig. 9. Fraction of the neutrons reaching the $^3$He detector after multiple scattering. The incident neutron energy is 0.4 MeV.**

*IV.C. Background subtraction*

A linear smooth non-resonant n-background was subtracted from below the resonance peaks. This procedure is justified by noting that all non-resonant samples of neighboring Z produced a similar smooth *non-peaked* background. The procedure is illustrated in Fig. 10 for the 423 keV case. A Gaussian plus a smooth linear background was fitted to the peak. Using the obtained parameters integration was performed in the energy range 380-460 keV yielding a background of 11195 and a net area of 4912, hence a statistical error of 2.6%. The peak at 99 keV is stronger with a relative smaller background yielding a statistical error of 1.2% (integration in the energy range 85-110 keV).

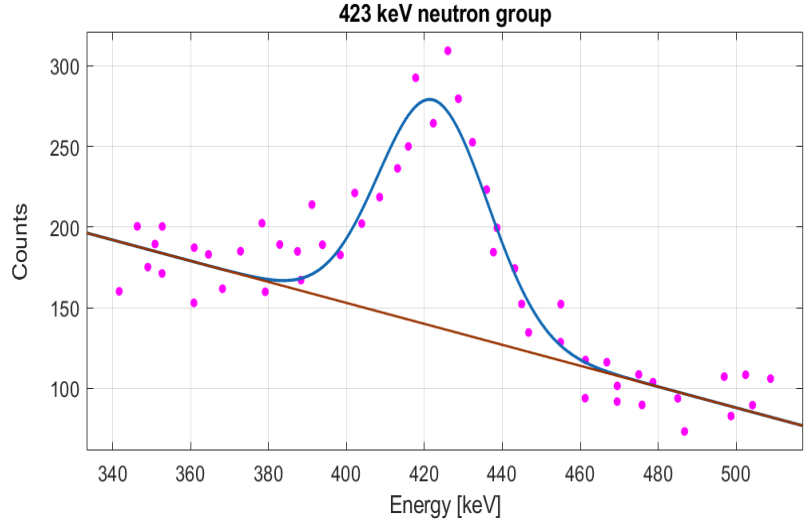

**Fig. 10. Background subtraction and analysis of the neutron group at 423 keV.**

In calculating the total error in the σ(γ,n) cross sections we accounted for the statistical errors, for the 6% errors in the relative efficiencies of the ³He detector (taken from Franz[21] at 120 keV and 460 keV, are 6% and 5.5% respectively), for the error in the reference cross section at 86 keV resonance (13.5%) and for the error in the integrated angular distribution (2.3% - obtained by drawing 1000 samples of its parameters from normal distributions and averaging). All of which were added quadratically. The statistical counting errors (i.e., those of the net areas) play a small role compared with the reference cross section error and the relative efficiency errors, resulting in a 16% error in both measured cross sections. Thus, the (γ,n) cross sections were found to be 35 ± 6 mb for the 423 keV resonance and 107 ± 17 mb for the 99 keV resonance.

### IV.D. Effect of Photon Contamination

The two neutron sources are contaminated with low energy photons of ~ 0.5 MeV produced by Compton scattering at 90° of the high intensity lines of the γ-sources on the high Z targets and by the 0.511 MeV photons obtained by positron annihilation. The shield of the ³He gas consists of 1.2 mm steel (the detector body), 2 mm lead and 0.5 mm cadmium. This shield is only partially effective against these photons, some ~36% transmission at 0.5 MeV.

Electrons can produce a signal in the detector when the counting rate is high and if there is a pileup. Electron-positron pairs are produced in the pair-production process, in the sample or in the scattering chamber walls, by photons with $E_\gamma > 1.022$ MeV, more abundantly by the higher energy γ rays ($E_\gamma$>5 MeV). In this case the shield is very effective, for an energetic electron of 3.5 MeV (i.e., from a gamma of ~8 MeV) the transmission coefficient is only 1.3% (calculated with EGS5[23]). Individual electrons cannot deposit more than ~300 keV in the $^3$He gas, and such pulses have amplitudes well below the thermal peak.

We now discuss the effect of photon contamination of the 99 keV neutron source. In addition to the low energy photons, this n-source is also contaminated with high energy photons. These photons are produced in a resonant process exactly like the neutrons. The sample nucleus is photoexcited and subsequently decays by emitting a photon. The $^{205}$Tl isotope is a strong resonance scatterer of the 7646 keV γ-line of the Fe(n,γ) reaction[22] with a γ-scattering cross section of 0.59 b. The effect of such photon contamination on the detection of the 99 keV neutrons is much smaller in comparison to that of the 86 keV n-source[13] which was strongly contaminated by the 7279 keV γ-line (emitted by the Fe(n,γ) source) and known to be resonantly scattered by $^{208}$Pb having a very large scattering cross section of 5.6 b. Since the resolution of the 86 keV neutron line, as measured by the $^3$He detector, was not affected[13] by such a strong intensity high energy photon contaminant, one would expect a negligible effect on the 99 keV neutron peak also.

The n- and $\gamma$-backgrounds of the 99 keV n-source are in principle higher than the 423 keV n-source. This is because the Fe(n,γ) source has strong intensity $\gamma$-lines at higher energies: 7279, 7632, 7646 and 9298 keV, producing higher neutron yields, being above the ($\gamma$,n) threshold for some stable isotopes of Tl. This causes an increase in the n- and hence of the $\gamma$-background via the (n,$\gamma$) reaction. However, this higher background was small relative to the strong intensity of 99 keV neutron peak created by the $^{205}$Tl($\gamma$,n) reaction.

## V. Increasing the intensity of the neutron sources

The intensity of the 99 keV n-source (produced by the Fe-$^{205}$Tl combination) can, in principle, be increased by a factor of ~ 300 i.e., to ~ $2 \times 10^7$ n/sec by using a larger mass (100 g) of the (γ,n) target of Tl with a larger diameter ~6.0 cm and employing a higher n-flux reactor of ~$10^{14}$ n/cm$^2$/s and if possible, a shorter distance of ~5 m between the γ-source and the (γ,n) target. The use of a larger diameter of the Tl target is to diminish the attenuation of γ-ray intensity within the target. In a similar manner, the intensity of the 423 keV n-source created by the V-$^{207}$Pb combination can be increased to ~$6 \times 10^5$ n/sec by using a larger isotopic $^{207}$Pb mass (of ~80 g) in conjunction with a higher n-flux reactor of $10^{14}$ n/cm$^2$/s.

It should be stressed that there is a very strong dependence on the type of the reactor available and its geometry. Not two reactors are the same and it should not to be expected to receive exactly the numbers of the present paper at other installations.

## VI. Conclusions

It was shown that two relatively strong and quasi monoenergetic neutron sources with energies of 99 keV and 423 keV can be produced, at a nuclear reactor, using a combination of (n,γ) and (γ,n) reactions. The method depends on a chance overlap between a discrete γ-line produced by n- capture on V or Fe and *isolated* resonance levels in $^{207}$Pb and $^{205}$Tl respectively.

The energy of these neutron sources, the first at ~100 keV and the second at ~400 keV, is of particular interest because each source is quasi monochromatic. They are quite different from the conventional n-sources (see a collection of measured PHS in Lorch[24]) such as that of $^{252}$Cf which is spread over a wide energy range of 10 eV to 10 MeV or the $^{241}$AmBe source covering the 100 eV to 10 MeV region.

Due to the complexity of the setup, there is little experience in using this kind of sources. In our group, which pioneered them, use was made mainly in the field of nuclear spectroscopy[13-16.] Their principal advantage is being quasi monochromatic, and as such, can, in principle, be used for energy and efficiency calibrations of n-detectors. Not an actual attempt was made until now in this direction. Determination of the absolute efficiency of a $^3$He spectrometer is a thorny one. Even in the work of Franz[21] et. al., which used a large number of monoenergetic neutron energies obtained with the $^7$Li(p,n) reaction, only the relative efficiency is given, and even this after renormalization to unpublished results of one of the original developers of the detector (Cuttler).

The intensity of the sources is comparable with the continuous α and γ induced sources but can not compete with the spallation or accelerator driven sources. Studies of $^7$Li(p,n) sources, to be used in BNCT – boron neutron capture therapy, quote[25] intensities of the order of $10^9$ n/sec, while, when the final stage of the SARAF[26] accelerator will be completed (using a liquid cooled Lithium source), similar intensity is expected, not necessarily for therapy.

**Appendix A**

*A.1 The angular distribution of the 423 keV n-group*

The measured angular distribution of the 423 keV neutrons was fitted to an expansion of even Legendre polynomials using only the first two terms:

$$W(\theta) = A_o + A_2 P_2(\cos\theta) \qquad (A1)$$

Where θ is the angle between the incident photon beam and the emitted neutron in the scattering plane; the deduced ratio was: $A_2/A_o = -0.538 \pm 0.055$. The symmetry of the distribution is important in defining the angular momenta of the emitted neutron and the spin of the emitting level. To check it, we carried out a goodness of fit test: the statistics is: $\chi^2 = 7.51$ versus a critical value of 9.49 at the customary 5% significance level. Therefore, a symmetric Legendre polynomial describes satisfactorily the measured data, with a p-value of about 11.1%, higher than the significance level of 5% (even higher than a more stringent significance level of 10%). Any asymmetry in the angular distribution may be caused by a mixture of two different multipolarities in the formation of the level but is unlikely to occur in the present case.

*A.2 Neutrons emitted from 7163 keV level of $^{207}$Pb*

The angular distribution (Fig. 8) corresponds to an emitted neutron with an angular momentum $l_n = 2$. Since the ground state of $^{207}$Pb is $J_0^\pi = 1/2^-$, denoting by $J$ the spin of the emitting resonance level in $^{207}$Pb, the conservation of angular momenta requires that: $\mathbf{J} + \mathbf{l_n} + \mathbf{s_n} = \mathbf{J_f}$, where $s_n$ is the neutron spin and $J_f$ is the ground state spin of the final nucleus. Hence, to conserve parity and angular momentum, it seems that the resonance level at 7163 keV in $^{207}$Pb has a spin and parity of $J^\pi = 3/2^+$ (being photoexcited by $E1$ transition) and the emitted neutron proceeds with $l_n = 2$ to the $^{206}$Pb ground state having $J_f^\pi = 0^+$.

## A.3 Neutrons emitted from the 7646 keV level of $^{205}$Tl

Since the ground state spin in $^{205}$Tl is **1/2$^+$** hence the 7646 keV level is very likely **$J^\pi$ =1/2$^-$** because of its large width which is characteristic of *E*1 transitions. It follows that the emitted neutrons proceed to the **$J_f^\pi$ = 2$^-$** ground state of $^{204}$Tl by a d-wave transition, **$l_n$ = 2** (see . Photoexcitation of the 7646 keV resonance level in $^{205Tl}$ showing the photo absorption process and the subsequent 99 keV neutron emission (with *ln* =2) leading to the $^{204Tl}$ ground state, $J_f^\pi$ = 2-, via the $^{205Tl}$(γ,n) reaction.

## A.4 Doppler Broadening

The cross section σ(γ,n) of the resonant process is a measure of the overlap between the incident γ-line energy and the nuclear level energy in the sample. This overlap is influenced by the thermal Doppler broadening of these two ingredients.

In the case of the 423 keV n-group, both the incident 7163 γ-line of the V(n,γ) reaction and the nuclear level in $^{207}$Pb are thermal Doppler broadened. The broadening is given by: $\Delta = E_\gamma(2kT_e/Mc^2)^{1/2}$, where $E_\gamma$ = 7163 keV, the energy of the incident line of the γ-source, k is the Boltzmann constant and M=51 is the vanadium atomic mass. $T_e$ is the effective temperature of metallic V, where the ambient temperature during reactor operation, is T= 550 K. To calculate $T_e$ for metallic V, we used the Lamb[27] formula, which requires a knowledge of the Debye temperature of Vanadium, $\Theta_D(V)$ = 390 K, from which we get $T_e$ = 560 K at T=550 K, yielding a spread of $\Delta_s$ = 10.2 eV, for the γ-line source. Similarly, the Doppler width of the resonance level in $^{207}$Pb is $\Delta_r$ = 3.8 eV; being smaller because M = 207 and $T_e$ = 300K (obtained using a Debye temperature, $\Theta_D(Pb)$ = 87 K), assuming a sample temperature of T= 298 K. Here however, this level is unbound having $\Gamma_n > 0$ with a total width $\Gamma = \Gamma_n + \Delta_r$. While these energy widths are essential for the overlap between the incident γ-line and the resonant level, both are smaller compared with the geometric energy spread calculated before, and not affecting the presented measurements.

In the same manner, the results of the Fe-Tl combination were calculated yielding: $\Delta_s$ = 11.5 eV (for the 7646 γ-line source) and $\Delta_r$ = 4.1 eV the Doppler width of the $^{205}$Tl nuclear level.

## A.5 Neutrons reaching the $^3$He detector through the shielding

It is of interest to quantify the influence of the detector shielding on the neutron penetration. This was estimated by Monte Carlo simulation, placing sheets of steel, lead and Cadmium (of appropriate thickness) and measuring the current behind these sheets onto a virtual surface 5x15 cm at 27 cm from them. The neutrons were emitted from a point source into a 6° opening, mimicking the angular opening of the detector at 90°. The results are 84.6%, 83.4%, 90.6% respectively, at 86, 99 and 423 keV.

**References**

[1] J. M. CARPENTER and W. B. YELON, "Neutron Scattering", in *Methods in Experimental Physics*, Vol. 23, Part A, Academic Press, Inc., pp. 99–196, (1986), https://doi.org/10.1016/S0076-695X(08)60555-4; A. FURRER, "Neutron Sources", in *Encyclopedia of Condensed Matter Physics*, Elsevier, 2005, pp. 69–75, https://doi.org/10.1016/b0-12-369401-9/00633-1; https://www.nrc.gov/docs/ML1122/ML11229A704.pdf (accessed Jan. 20, 2022).

[2] K. H. BECKURTS and K. WIRTZ, "Neutron Physics" Springer-Verlag, New York, (1964).

[3] A. N. GARG and R. J. BATRA, "Isotopic sources in neutron activation analysis", *J. Radioanalytical Nucl. Chem.*, **98,** 167 (1986); https://doi.org/10.1007/BF02060444

[4] https://raims.co.uk/product/americium-241-beryllium-ambe-sealed-sources (accessed Jan. 20, 2022)

[5] K. A. ALFIERI, R. C. BLOCK and R. J. TURINSKY, "Measurement and Evaluation of Total Neutron Cross-Section Minima in Elemental Iron from 24 to 750 keV", *Nucl. Sci. Eng.*, **51,** 25 (1973); https://dx.doi.org/10.13182/NSE73-A23254.

[6] R. MOREH, R.C. BLOCK and Y. DANON, "Generating a multi-line neutron beam using an electron Linac and a U-filter", *Nucl. Inst. Meth. A*, **562**, 401 (2006); https://doi.org/10.1016/j.nima.2006.02.160.

[7] A. MICHAUDON, S. CIERJACK and R. E. CHRIEN, "Neutron Sources for Basic Physics and Application", Pergamon press, Oxford, (1983); G. D. KIM et. al., "Mono-energetic Neutron Source Using the $^7$Li(p,n)$^7$Be Reaction", *J. Korean Phys. Soc*. **55,** 1404 (2009); https://doi.org/10.3938/jkps.55.1404.

[8] J. W. G. THOMASON, "The ISIS Spallation Neutron and Muon Source—The first thirty-three years", *Nucl. Inst. Meth*. A, **917,** 61 (2019); https://doi.org/10.1016/j.nima.2018.11.129.

[9] https://neutrons.ornl.gov/sns "Spallation Neutron Source" (accessed January 20, 2022); A detailed and comprehensive reference to spallation is: D. FILGES and F. GOLDENBAUM, "Handbook of Spallation Research". Wiley-VCH Verlag GmbH & Co. (2009); https://doi.org/10.1002/9783527628865.

[10] https://ntof-exp.web.cern.ch "n-TOF The neutron time-of-flight facility at CERN" (accessed January 20, 2022).

[11] C. GUERRERO et al., "Performance of the neutron time-of-flight facility n-TOF at CERN", *Eur. Phys. J. A,* **49,** 27 (2013); https://doi.org/10.1140/epja/i2013-13027-6.

[12] G. L. MOLNAR et. al., "The new prompt gamma-ray catalogue for PGAA", *Applied Radiation and Isotopes*, **53,** 527 (2000); https://doi.org/10.1016/S0969-8043(00)00209-8.

[13] R. MOREH, Y. BIRENBAUM and Z. BERANT, "A new 86 keV neutron source from the $^{207}$Pb(γ, n) reaction", *Nucl. Inst. Meth.*, **155,** 429 (1978); https://doi.org/10.1016/0029-554X(78)90525-6

[14] R. MOREH et. al., "A strong dynamic neutron source based on the $^{209}$Bi(γ, n) reaction", *Nucl. Inst. Meth. A,* **309,** 503 (1991); https://doi.org/10.1016/0168-9002(91)90255-O

[15] Z. BERANT et. al., "Study of the $^{49}$Ti energy levels using the (γ, γ′) and the (γ, n) reactions", *Nucl. Phys. A,* **368,** 201 (1981); https://doi.org/10.1016/0375-9474(81)90682-5

[16] Y. BIRENBAUM et. al., "Angular distributions of photoneutrons from the $^{207,\,208}$Pb(γ, n$_0$) reactions", *Nucl. Phys. A,* **369,** 483 (1981); https://doi.org/10.1016/0375-9474(81)90033-6

[17]  https://www.nndc.bnl.gov/capgam/byTarget/z023_51V.html (accessed Jan. 21, 2022)

[18] S. SHALEV and J. CUTTLER, "The Energy Distribution of Delayed Fission Neutrons", *Nucl. Sci. Eng.,* **51,** 52 (1973); https://doi.org/10.13182/NSE73-A23257

[19] M. WANG et. al., "The AME2016 atomic mass evaluation (II). Tables, graphs and references", *Chinese Physics C,* **41,** 030003 (2017); https://doi.org/10.1088/1674-1137/41/3/030003.

[20] D. J. HOREN, J. A. HARVEY and N. W. HILL, "Doorway states in s-, p-, and d-wave entrance channels in $^{207}$Pb+n reaction", *Phys. Rev. C* **18,** 722 (1978); https://doi.org/10.1103/PhysRevC.18.722.

[21] H. Franz et. al., "Delayed-neutron spectroscopy with $^3$He spectrometers", *Nucl. Inst. Meth.*, **144,** 253 (1977); https://doi.org/10.1016/0029-554X(77)90116-1.

[22] R. MOREH and A. WOLF, "Study of the Energy Levels of $^{205}$Tl Using the (γ,γ′) Reaction", *Phys. Rev.,* **182,** 1236 (1969); https://doi.org/10.1103/PhysRev.182.1236.

[23] H. HIRAYAMA et. al., High Energy Accelerator Research Organization (KEK) Japan, SLAC-R-730/KEK-2005-8 Report, (August 2005).

[24] E. A. LORCH, "Neutron Spectra of $^{214}$Am/B, $^{241}$Am/Be, $^{241}$Am/F, $^{242}$Cm/Be, $^{238}$Pu/$^{13}$C and $^{252}$Cf isotopic neutron sources", *Intern. J. Appl., Rad. Isot.*, **24,** 585 (1973); https://doi.org/10.1016/0020-708X(73)90127-0.

[25] R. G. FAIRCHILD et. al., "Installation and Testing of an Optimized Epithermal Neutron Beam at BMRR", in *Neutron Beam Design, Development, and Performance for Neutron Capture Therapy*, pp. 185-199, Edited by O. K. Harling et al, Plenum Press, New York, (1990); https://doi.org/10.1007/978-1-4684-5802-2_14.

[26] M. Friedman et. al., "Simulation of the neutron spectrum from the $^7$Li(n,p) reaction with a liquid-lithium target at SARAF", *Nucl. Inst. Meth. A*, **698,** 117 (2013); https://doi.org/10.1016/j.nima.2012.09.027.

[27] W. E. Lamb, Jr., "Capture of Neutrons by Atoms in a Crystal", *Phys. Rev.*, **55** 190, (1939); https://doi.org/10.1103/PhysRev.55.190.